\documentclass[amsmath,amssymb,pra,aps,showpacs,superscriptaddress,twocolumn]{revtex4-1}
\usepackage{amsmath,amsfonts,amssymb,amsthm,graphics,graphicx,epsfig,bbm}
\usepackage[colorlinks=true,citecolor=blue,linkcolor=blue,urlcolor=blue]{hyperref}
\usepackage[usenames]{color}
\usepackage{graphicx}
\usepackage{subfigure}
\usepackage{amsmath}
\usepackage{epsfig}
\usepackage{dcolumn}
\usepackage{bm}
\usepackage{color}
\usepackage{times}
\usepackage{epstopdf}
\usepackage{amssymb}
\usepackage{amstext}
\usepackage{latexsym}
\usepackage{hyperref}
\usepackage{amsfonts}
\usepackage{psfrag}
\usepackage{soul,xcolor}
\usepackage[normalem]{ulem}
\usepackage{dsfont}
\usepackage{txfonts}
\usepackage{mathtools}

\newcommand{\ket}[1]{\vert #1 \rangle}
\newcommand{\bra}[1]{\langle #1 \vert}

\begin{document}
\setstcolor{red}

\title{Spin-1 models in the ultrastrong coupling regime of circuit QED} 
\date{\today}

\author{F. Albarr\'an-Arriagada}
\email[F. Albarr\'an-Arriagada]{\qquad francisco.albarran@usach.cl}
\affiliation{Departamento de F\'isica, Universidad de Santiago de Chile (USACH), 
Avenida Ecuador 3493, 9170124, Santiago, Chile}

\author{L. Lamata}
\affiliation{Department of Physical Chemistry, University of the Basque Country UPV/EHU, Apartado 644, 48080 Bilbao, Spain}

\author{E. Solano}
\affiliation{Department of Physical Chemistry, University of the Basque Country UPV/EHU, Apartado 644, 48080 Bilbao, Spain}
\affiliation{IKERBASQUE, Basque Foundation for Science, Maria Diaz de Haro 3, 48013 Bilbao, Spain}
\affiliation{Department of Physics, Shanghai University, 200444 Shanghai, China}

\author{G. Romero}
\affiliation{Departamento de F\'isica, Universidad de Santiago de Chile (USACH), 
	Avenida Ecuador 3493, 9170124, Santiago, Chile}

\author{J. C. Retamal}
\affiliation{Departamento de F\'isica, Universidad de Santiago de Chile (USACH), 
	Avenida Ecuador 3493, 9170124, Santiago, Chile}
\affiliation{Center for the Development of Nanoscience and Nanotechnology 9170124, Estaci\'on Central, Santiago, Chile}

\begin{abstract}
We propose a superconducting circuit platform for simulating spin-1 models. To this purpose we consider a chain of $N$ ultrastrongly coupled qubit-resonator systems interacting through a grounded SQUID. The anharmonic spectrum of the qubit-resonator system and the selection rules imposed by the global parity symmetry allow us to activate well controlled two-body quantum gates via AC-pulses applied to the SQUID. We show that our proposal has the same simulation time for any number of spin-1 interacting particles. This scheme may be implemented within the state-of-the-art circuit QED in the ultrastrong coupling regime. 
\end{abstract}
\maketitle


\section{Introduction}
A major challenge in quantum physics is the development of capabilities to study dynamical properties of quantum many-body systems \cite{DeChiara2006,Heyl2013,Eisert2015,Hauke2016,Jurcevic2017,Ho2017}. Among these problems we have the study of interacting magnetic particles described by the Ising or Heisenberg model, possibly including anisotropy, which becomes intractable as the system size increases. In particular, spin-1 systems have drawn increasing interest due to the presence of diverse exotic phenomena such as the Haldane phase \cite{Haldane1983,Kennedy1992,Itoi1997,Pollmann2010} and quantum phase transitions \cite{Chen1973,Kolezhuk1996,Mila2000}.  Recent proposals have explored platforms to engineer interactions of effective spin-1 particles in order to study symmetry protected topological phases \cite{Cohen2014,Senko2015,Choi2017}. Further, finite size magnetic systems could be of great relevance since their properties depend on the system structure \cite{Lounis2008,Guidi2015,Romming2015,Lima2017,Wiesendanger2016}. Nevertheless, the exact calculation of dynamical properties of such systems is not possible using classical tools \cite{Lloyd1996} because the resources required for data storage scale exponentially with the number of particles in the system. Overcoming this problem requires a quantum simulator (QS) \cite{Georgescu2014} that, as conjectured by Feynman, needs only a data storage and processing capability that increases polynomially with the number of particles \cite{Feynman1982}.  

A highly scalable and tunable technology for QSs is the superconducting circuit architecture \cite{Blais2004,Mezzacapo2014,Lamata2017,LasHeras2014,Barends2015,Salathe2015}. This technology allows for the study of light-matter interaction in the ultrastrong (USC) \cite{Niemczyk2010,FornDiaz2010,FornDiaz2017} and deep-strong coupling (DSC) regimes \cite{Casanova2010,Yoshihara2017}. The USC regime offers features such as anharmonic energy spectrum and parity symmetry, which lead to interesting theoretical applications such as fast quantum gates~\cite{Romero2012}, as well as high fidelity quantum state transfer \cite{Kyaw2017,CardenasLopez2017}, among others \cite{Ashhab2010,Stassi2013,Benenti2014,AlbarranArriagada2017}.

In this article, we propose to simulate a spin-1 chain of the Heisenberg and Ising models using ultrastrongly coupled light-matter systems, whose lowest three energy levels simulate the spin-1 particles. We can implement two-body interactions through coupling between resonators by means of grounded superconducting quantum interference devices (SQUIDs)~\cite{BookSquid}. We use two different interleaved qubit-resonator systems in order to simulate all nearest-neighbor interactions in a single gate, thus producing the same simulation time for any number of spin-1 interacting particles.

\begin{figure}[t]
	\centering
	\includegraphics[width=1\linewidth]{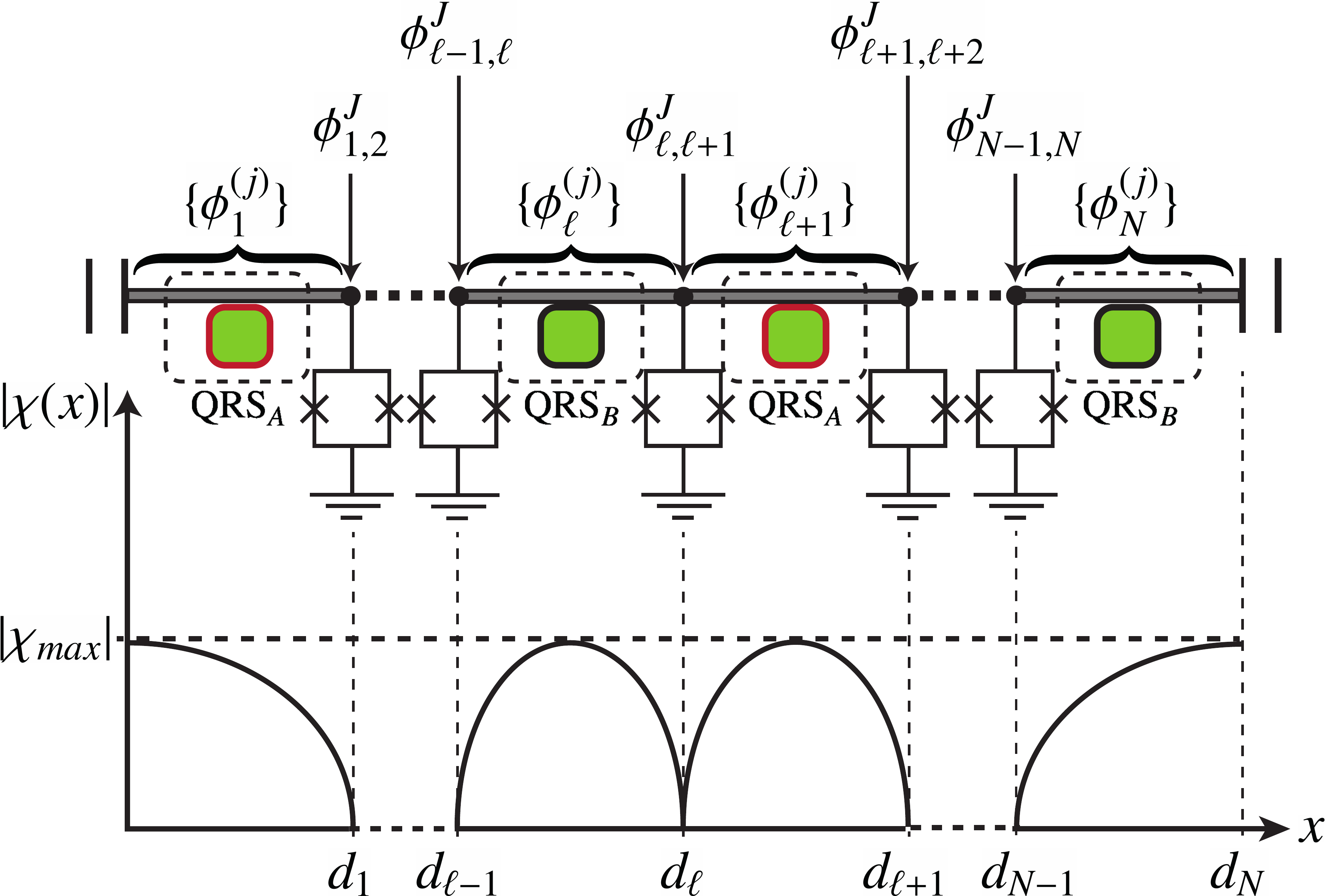}
	\caption{(color online). On top, chain of quantum Rabi systems, given by transmon-qubit (green box) ultrastrongly coupled to transmission lines (grey solid line). The QRS$_j$ and QRS$_{j+1}$ are coupled through grounded SQUID$_{j,j+1}$ (box with crosses). The chain is composed by interleaved species of QRSs (A and B). On bottom the spatial function $\chi(x)$ of the voltage, that define a $\lambda/4$-resonator for the edges, and $\lambda/2$-resonators for the bulk.}
	\label{Fig1}
\end{figure}
\section{The model}
Let us consider a chain of $N$ ultrastrongly coupled qubit-resonator systems, referred to as quantum Rabi systems (QRS), coupled by a grounded SQUID through their respective resonators ~\cite{Felicetti2014,Wang2016}, see Fig.~\ref{Fig1}. We stress that this system may be implemented in a circuit quantum electrodynamics platform where each QRS is built of a superconducting resonator coupled to a transmon qubit \cite{Kraglund2017,Bosman2017}. The transmons must be located at the edges of the resonators for the outer QRSs of the chain and at the center of each resonator for the remaining QRSs as is shown in Fig.~\ref{Fig1}. This ensures a maximum coupling between the transmon and the electric field distribution over each resonator, which is imposed by zero voltage boundary conditions at the SQUIDs. The Hamiltonian that describes this system reads (see appendix \ref{AppendixA})
\begin{eqnarray}
&&H=\sum_{\ell=1}^{N}\bigg[H^{\textrm{QRS}}_{\ell}+\bigg(P_{\ell}^{\ell,\ell+1}+P_{\ell}^{\ell-1,\ell}\bigg)(a^{\dagger}_{\ell}+a_{\ell})^2\bigg]\nonumber\\
&&-\sum_{\ell=1}^{N-1}\bigg[2\sqrt{P_{\ell}^{\ell,\ell+1} P_{\ell+1}^{\ell,\ell+1}}(a_{\ell}^{\dagger}+a_{\ell})(a_{\ell+1}^{\dagger}+a_{\ell+1})\bigg]\nonumber\\
&&+\sum_{\ell=1}^{N}\bigg[\bigg(Q_{\ell}^{\ell,\ell+1}\bar{\Phi}_{\ell,\ell+1}(t)+Q_{\ell}^{\ell,\ell-1}\bar{\Phi}_{\ell,\ell-1}(t)\bigg)(a^{\dagger}_{\ell}+a_{\ell})^2\bigg]\nonumber\\
&&-\sum_{\ell=1}^{N-1}\bigg[2\sqrt{Q_{\ell}^{\ell,\ell+1} Q_{\ell+1}^{\ell,\ell+1}}\bar{\Phi}_{\ell,\ell+1}(t)(a_{\ell}^{\dagger}+a_{\ell})(a_{\ell+1}^{\dagger}+a_{\ell+1})\bigg],
\label{Eq.01}
\end{eqnarray}
where $\ell$ refers to the $\ell$th site of the chain, the pair $(\ell,\ell+1)$ refers to the SQUID$_{\ell,\ell+1}$ between the sites $\ell$ and $\ell+1$, $\bar{\Phi}_{\ell,\ell+1}(t)$ is the external time-dependent magnetic flux threading the SQUID$_{\ell,\ell+1}$, $a_{\ell}(a_{\ell}^{\dagger})$ is the annihilation (creation) operator of the lowest mode of the resonator, $P_{\ell}^{\ell,\ell+1}$ and $Q_{\ell}^{\ell,\ell+1}$ are time-independent constants given by
\begin{eqnarray}
&&P_{\ell}^{l,l+1}=\frac{\varphi_o\omega_{\ell}^r}{4I_c Z_{\ell}^{2}C_{\ell}}\frac{1}{\cos(\bar{\Phi}_o^{l,l+1})},\nonumber\\
&&Q_{\ell}^{l,l+1}=\frac{\varphi_o\omega_{\ell}^r}{4I_c Z^{2}_{\ell}C_{\ell}}\frac{\sin(\bar{\Phi}_o^{l,l+1})}{\cos^2(\bar{\Phi}_o^{l,l+1})},
\label{Eq.02}
\end{eqnarray}
with $\varphi_o$ is flux quantum, $I_c$ the critical current; $\omega_{\ell}^r$, $C_{\ell}$ and $Z_{\ell}$ are the fundamental frequency, capacitance, and impedance, respectively, that characterize the $\ell$th site. $\bar{\Phi}_o^{l,l+1}$ is the offset component of the external magnetic flux threading the SQUID$_{l,l+1}$. Additionally, we use two interleaved species of QRS, in order to obtain controllable two-body interactions, as will be shown in the next section. Finally, 
$H^{\textrm{QRS}}_{\ell}$ is the Hamiltonian of the $\ell$th QRS described by the quantum Rabi model~\cite{Rabi1937,Braak2011}
\begin{eqnarray}
H^{\textrm{QRS}}_{\ell}=\frac{\hbar \omega^{q}_{\ell}}{2}\sigma_{\ell}^{z}+\hbar\omega^r_{\ell}a_{\ell}^{\dagger}a_{\ell}+\hbar g_{\ell}\sigma_{\ell}^x\left(a_{\ell}^{\dagger}+a_{\ell}\right),  
\label{Eq.03}
\end{eqnarray}
with  $\sigma_{\ell}^k$ is the $k$-Pauli matrix associated with the qubit of the QRS. In addition, $\omega^{q}_{\ell}$ is the qubit frequency  and $g_{\ell}$ the qubit-resonator coupling strength. The diagonalization of Eq. (\ref{Eq.03}) defines the eigenbasis $\{\ket{j}_{\ell}\}$ as
\begin{equation}
H^{\textrm{QRS}}_{\ell}\ket{j}_{\ell}=\lambda_j^{\ell}\ket{j}_{\ell}
\label{Eq.04}
\end{equation}
where $j=\{0,1,2,...,\infty\}$, and  $\lambda_j^{\ell}$ is the eigenenergy of the $j$th eigenstate $\ket{j}_{\ell}$ of the $\ell$th QRS. The spectrum of the QRS is anharmonic and exhibits parity symmetry defined by the operator $\Pi_{\ell}=e^{i\pi(a^{\dag}_{\ell}a_{\ell}+\sigma^{+}_\ell\sigma^{-}_\ell)}$~\cite{Braak2011,Albert2012,Wolf2012}. These properties allow us to engineer a spin-1
particle with the three lowest energy levels of a QRS.
\begin{figure}[t]
	\centering
	\includegraphics[width=0.8\linewidth]{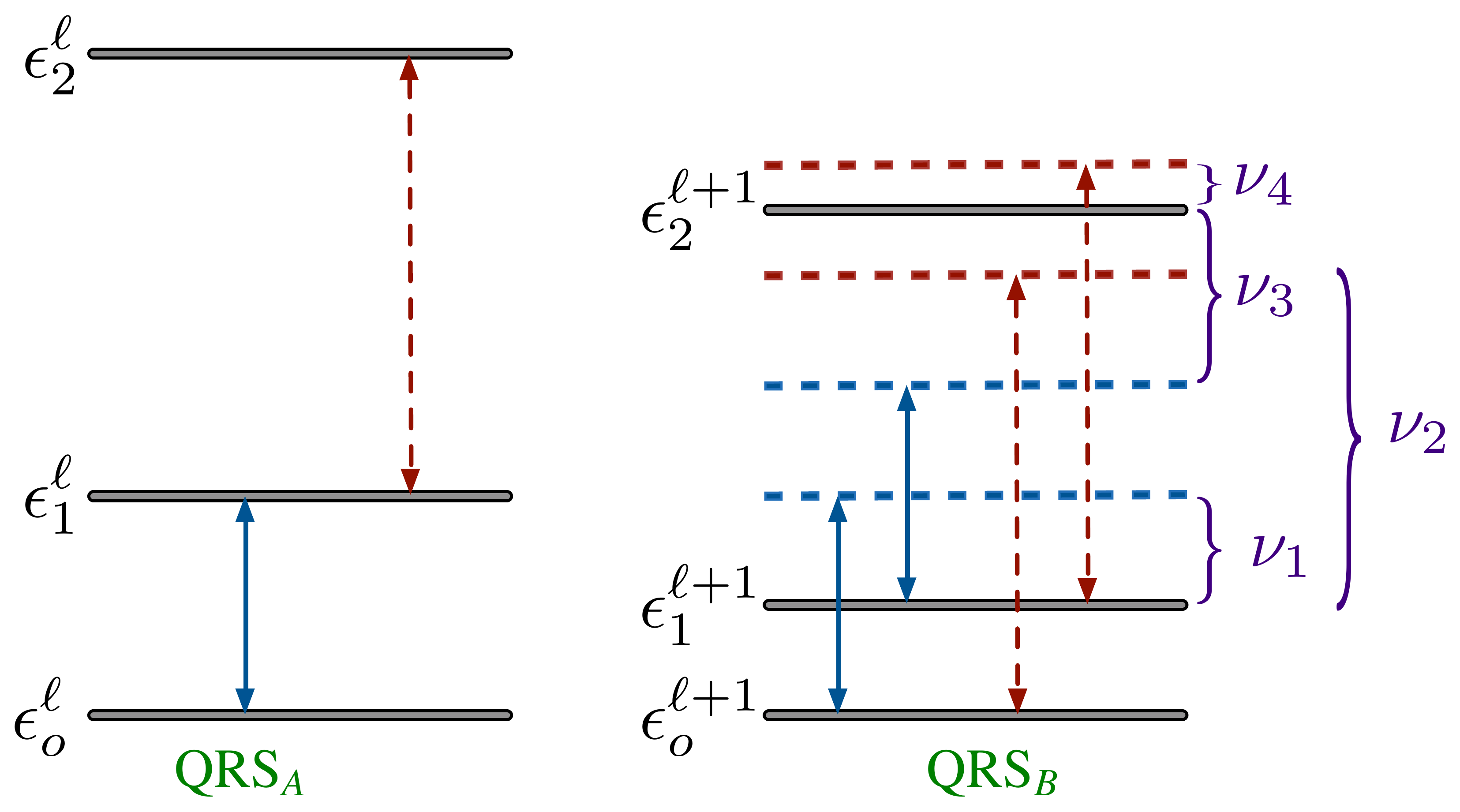}
	\caption{Energy-level diagram for two adjacent QRSs. The blue solid arrows match the $\epsilon^{\ell}_1-\epsilon^{\ell}_o$ transition, and the red dashed arrows match the $\epsilon^{\ell}_2-\epsilon^{\ell}_1$ transition. $\nu_n$ are the necessary frequencies to adjust the resonance condition for different transitions.}
	\label{Fig2}
\end{figure}
\section{Two-body interactions}
\begin{figure}[t]
	\centering
	\includegraphics[width=0.9\linewidth]{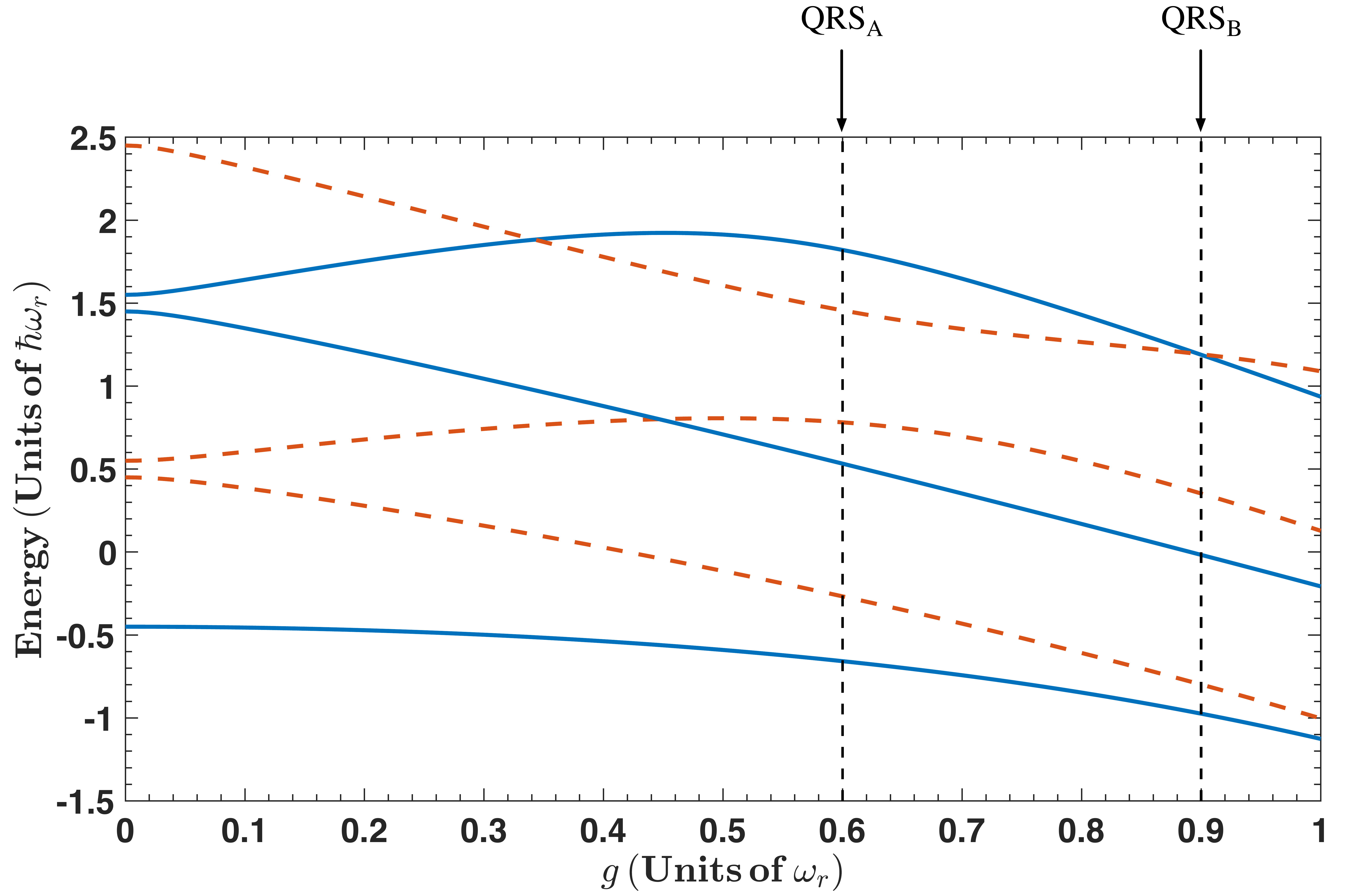}
	\caption{Energy diagram of Hamiltonian (\ref{Eq.07}). Blue continuous line indicate the states with parity $+1$ and red dashed lines states with parity $-1$, the vertical dashed lines indicate the coupling strength for the QRS$_A$ and for the QRS$_B$ used in our numerical calculations.}
	\label{Fig3N}
\end{figure}

Let us focus on two adjacent QRSs for arbitrary sites $\ell$ and $\ell+1$, these QRS have different energy spectrum (see Fig. \ref{Fig2} and Fig.~\ref{Fig3N}). This means that the chain will be composed by two interleaved species of QRSs (A and B) as is shown in Fig.~\ref{Fig1}, such that each QRS has a different spectrum compared with its nearest-neighbours. To view the effect of this condition in the quantum dynamics, it is instructive to see which terms of Hamiltonian (\ref{Eq.01}) will play a role in the implementation of two-body interactions. 

Before we continue our discussion, it is instructive to write the field operator $(a_{\ell}^{\dagger}+a_{\ell})$ in the basis $\{\ket{j}_{\ell}\}$ defined by Eq. (\ref{Eq.04}) as
\begin{equation}
a_{\ell}+a^{\dag}_{\ell}=\sum_{j,k>j=0}\chi^{\ell}_{k,j}\ket{k}_\ell\bra{j}+{\rm H.C},
\label{Eq.05}
\end{equation}
where $\chi^{\ell}_{k,j}={_\ell}\bra{k}(a_{\ell}+a_{\ell}^{\dagger})\ket{j}_{\ell}$. This operator can only relate states with different parity, thus  $\chi^{\ell}_{k,j}=0$ when  $\ket{k}_{\ell}$ and $\ket{j}_{\ell}$ have the same parity. Now, the first interaction operator that we consider is the single-mode squeezing term $(a_{\ell}^{\dagger}+a_{\ell})^2$, this is a parity preserving operator over the subspace defined by $H_{\ell}^{\textrm{QRS}}$.~\cite{Wang2016}. This means that
they can only produce transitions between states of equal parity and energy shifts for a single QRS according to 
\begin{equation}
(a_{\ell}+a^{\dag}_{\ell})^2=\sum_{j,k>j=0}[z^{\ell}_{kj}\ket{k}_\ell\bra{j}+{\rm H.C}]+\sum_{j}z^{\ell}_{jj}\ket{j}_\ell\bra{j},
\label{Eq.06}
\end{equation}
where $z^{\ell}_{kj}= {_\ell}\bra{k}(a_{\ell}+a^{\dag}_{\ell})^2\ket{j}_\ell=\sum_{l=1}^{\infty}\chi^{\ell}_{kl}\chi^{\ell}_{lj}$ are matrix elements in the eigenbasis $\{\ket{j}_{\ell}\}$. The last term of Eq.~(\ref{Eq.06}) together with $H^{\textrm{QRS}}_{\ell}$ will define the diagonal time-independent Hamiltonian for the $\ell$th site as
\begin{equation}
	H_o^{\ell}=\sum^{\infty}_{j=0}\bigg[\lambda_j^{\ell}+\bigg(P_\ell^{\ell,\ell-1}+P_\ell^{\ell-1,\ell}\bigg)z_{jj}^{\ell}\bigg]\ket{j}_\ell\bra{j}=\sum^{\infty}_{j=0}\epsilon_j^{\ell}\ket{j}_\ell\bra{j}.
	\label{Eq.07}
\end{equation}

Second, we consider the interacting terms $(a_{\ell}^{\dagger}+a_{\ell})(a_{\ell+1}^{\dagger}+a_{\ell+1})$ in Eq.~(\ref{Eq.01}), they appear as time-independent (second line) and time-dependent (fourth line) contributions that might lead to the desired two-body interactions between different species. These operators preserve the global parity of the system, but change the local parity of the pair of QRS involved in the interaction according to Eq.(\ref{Eq.05}).

Since the QRS$_{\ell}$ has different energy spectrum compared to QRS$_{\ell+1}$, the time-independent interactions (second line of Eq. (\ref{Eq.01})) are off-resonant and can be neglected by applying a rotating wave approximation (RWA) with respect to $H_o=\sum_{\ell}H_o^{\ell}$ for a specific choice of system parameters. Under similar condition one could neglect single-body transitions induced by matrix elements $z_{jk}^{\ell}$, in Eq. (\ref{Eq.06}). Finally in this way, the last terms in Eq. (\ref{Eq.01}) might implement two-body interactions for a proper choice of resonant condition in the magnetic flux $\bar{\Phi}_{\ell,\ell+1}$, as we will demonstrate below.

The magnetic flux $\bar{\Phi}_{\ell,\ell+1}$ can be written for all SQUIDs as a linear superposition of harmonic signals
\begin{equation}
\bar{\Phi}_{\ell,\ell+1}=\sum_{n}\gamma_{n}\cos(\nu_{n}t),
\label{Eq.08}
\end{equation}
where $\gamma_{n}$ and $\nu_{n}$ are the amplitude and frequency of $n$th component respectively. To see how effective two-body interactions are achieved by using this magnetic signal, we write the Hamiltonian (\ref{Eq.01}) in the interaction picture with respect to $H_o$ for adjacent sites $\ell$ and $\ell+1$
\begin{eqnarray}
&&H_I^{\ell,\ell+1}=\sum\limits_{j,k>j}\bigg[\bigg(P_{\ell}^{\ell,\ell+1}+P_{\ell}^{\ell-1,\ell}\bigg)z_{kj}^{\ell}e^{i(\epsilon_k^{\ell}-\epsilon_j^{\ell})t}\ket{k}_{\ell}\bra{j}\bigg]\nonumber\\
&&-2\sqrt{P_{\ell}^{\ell,\ell+1}P_{\ell+1}^{\ell,\ell+1}}\sum_{j,k>j}\sum_{l,m>l}\chi_{kj}^{\ell}\chi_{ml}^{\ell+1}\bigg(e^{i\delta_{kj}^{ml}t}\ket{k}_{\ell}\bra{j}\nonumber\\
&&+e^{i\Delta_{kj}^{ml}t}\ket{j}_{\ell}\bra{k}\bigg)\ket{m}_{\ell+1}\bra{l}+\sum_n\bigg[\sum\limits_{j,k>j}\frac{\gamma_n}{2}z_{kj}^{\ell}\bigg(Q_{\ell}^{\ell,\ell+1}+Q_{\ell}^{\ell-1,\ell}\bigg)\nonumber\\
&&\times\bigg(e^{i(\epsilon_k^{\ell}-\epsilon_j^{\ell}-\nu_n)t}+e^{i(\epsilon_k^{\ell}-\epsilon_j^{\ell}+\nu_n)t}\bigg)\ket{k}_{\ell}\bra{j}+\sum_{j}\frac{\gamma_n}{2}z_{jj}^{\ell}\bigg(e^{i\nu_n t}+e^{-i\nu_n t}\bigg)\nonumber\\
&&\times\bigg(Q_{\ell}^{\ell,\ell+1}+Q_{\ell}^{\ell-1,\ell}\bigg)\ket{j}_{\ell}\bra{j}\bigg]\nonumber\\
&&-\sum_{j,k>j}\sum_{l,m>l}\gamma_n\sqrt{Q_{\ell}^{\ell+1}Q_{\ell+1}^{\ell,\ell+1}}\chi_{kj}^{\ell}\chi_{ml}^{\ell+1}\bigg(\bigg(e^{i(\delta_{kj}^{ml}-\nu_n)t}+e^{i(\delta_{kj}^{ml}+\nu_n)t}\bigg)\ket{k}_{\ell}\bra{j}\nonumber\\
&&+\bigg(e^{i(\Delta_{kj}^{ml}-\nu_n)t}+e^{i(\Delta_{kj}^{ml}+\nu_n)t}\bigg)\ket{j}_{\ell}\bra{k}\bigg)\ket{m}_{\ell+1}\bra{l}+ {\rm H.C},
\label{Eq.09}
\end{eqnarray}
where $\delta_{kj}^{ml}=(\epsilon_m^{\ell+1}-\epsilon_l^{\ell+1})+(\epsilon_k^{\ell}-\epsilon_j^{\ell})$, and $\Delta_{kj}^{ml}=|(\epsilon_m^{\ell+1}-\epsilon_l^{\ell+1})-(\epsilon_k^{\ell}-\epsilon_j^{\ell})|$. Since we use two different interleaved species of QRSs ($A$ and $B$), all $\delta_{kj}^{ml}$ and $\Delta_{kj}^{ml}$ do not depend on $\ell$. Now, to obtain an effective two-body interaction we need to adjust a frequency $\nu_{n}$ to a specific transition, for example, if the magnetic flux (\ref{Eq.08}) has a component with frequency $\nu_{n}=\Delta_{10}^{10}$, the term proportional to $\ket{1}_\ell\bra{0}\otimes\ket{0}_\ell\bra{1} + \rm{H.c.}$ becomes resonant and will survive under the RWA with respect to $H_o$. In the same way, the term $\ket{1}_\ell\bra{0}\otimes\ket{1}_\ell\bra{0} + \rm{H.c.}$ becomes resonant if the frequency $\nu_n=\delta^{10}_{10}$. Then, for a proper choice of each harmonic component in Eq. (\ref{Eq.08}), one can
activate different transitions in Eq. (\ref{Eq.09}). An operator like
\begin{equation}
S_+^{\ell}S_-^{\ell+1}+S_-^{\ell}S_+^{\ell+1}=\sum_{j=0}^{1}\ket{j}_{\ell}\bra{j+1}\otimes\ket{j+1}_{\ell+1}\bra{j}+\textrm{H.C},
\label{Eq.10}
\end{equation}
can be engineered by setting $\bar{\Phi}(t)$ to be composed of four signals with frequencies $\nu_1=\Delta_{10}^{10}$, $\nu_2=\Delta_{10}^{21}$, $\nu_3=\Delta_{21}^{10}$, $\nu_4=\Delta_{21}^{21}$, and amplitudes $\gamma_1=f/(\chi_{10}^{\ell}\chi_{10}^{\ell+1})$, $\gamma_2=f/(\chi_{10}^{\ell}\chi_{21}^{\ell+1})$, $\gamma_3=f/(\chi_{21}^{\ell}\chi_{10}^{\ell+1})$, $\gamma_4=f/(\chi_{21}^{\ell}\chi_{21}^{\ell+1})$, $\gamma_{n>4}=0$, where $f$ is a manipulable parameter proportional to the amplitude of the magnetic flux threading the SQUID. For $\mathcal{C}_{xy}^{\ell,\ell+1}=f\sqrt{Q_{\ell}^{\ell,\ell+1}Q_{\ell+1}^{\ell,\ell+1}}\ll\Delta_{kj}^{ml}$, we can apply the RWA with respect to $H_o$ to obtain the effective Hamiltonian
\begin{eqnarray}
H_{XY}^{(\ell,\ell+1)}(\mathcal{C}_{xy}^{\ell,\ell+1})&&=C_{xy}^{\ell,\ell+1}\bigg(S_X^{\ell}S_X^{\ell+1}+S_Y^{\ell}S_Y^{\ell+1}\bigg)\nonumber\\
=&&C_{xy}^{\ell,\ell+1}\bigg[\ket{1}_{\ell}\bra{0}\otimes\bigg(\ket{0}_{\ell+1}\bra{1}+\ket{1}_{\ell+1}\bra{2}\bigg)\nonumber\\
&&+\ket{2}_{\ell}\bra{1}\otimes\bigg(\ket{0}_{\ell+1}\bra{1}+\ket{1}_{\ell+1}\bra{2}\bigg)\bigg]+{\rm H.C,}
\label{Eq.11}
\end{eqnarray}
where 
\begin{eqnarray}
&&S_X^{\ell}=\frac{1}{\sqrt{2}}\bigg(\ket{1}_{\ell}\bra{0}+\ket{2}_{\ell}\bra{1}+\ket{0}_{\ell}\bra{1}+\ket{1}_{\ell}\bra{2}\bigg),\nonumber\\ &&S_Y^{\ell}=\frac{1}{\sqrt{2}}\bigg(-i\ket{1}_{\ell}\bra{0}-i\ket{2}_{\ell}\bra{1}+i\ket{0}_{\ell}\bra{1}+i\ket{1}_{\ell}\bra{2}\bigg).
\label{Eq.12}
\end{eqnarray}
Furthermore, since each SQUID is connecting one QRS$_A$ and a QRS$_B$, we can set $\mathcal{C}_{xy}^{\ell,\ell+1}=\mathcal{C}_{xy}=f\sqrt{Q_AQ_B}$ for all $\ell$. Figure~\ref{Fig2} shows the energy diagram of both QRSs and the frequencies $\nu_n$ needed to obtain Eq.~(\ref{Eq.09}). Since all magnetic fluxes $\bar{\Phi}_{\ell,\ell+1}$ are independent, they can be switched on at the same time, so the Hamiltonian (\ref{Eq.09}) can be simulated for the entire chain given by
\begin{equation}
\bar{H}_{XY}=\sum_{\ell=1}^{N-1}H_{xy}^{\ell,\ell+1}(\mathcal{C}_{xy})=\mathcal{C}_{xy}\sum_{\ell=1}^{N-1}\Bigg(S_X^{\ell}S_X^{\ell+1}+S_Y^{\ell}S_Y^{\ell+1}\Bigg).
\label{Eq.13}
\end{equation}

To complete the simulation of the Heisenberg model for adjacent spin-1 particles we need to simulate the following term
\begin{eqnarray}
S_X^{\ell}S_X^{\ell+1}=&&\sum_{j=0}^{1}\bigg[\ket{j}_{\ell}\bra{j+1}\otimes\ket{j}_{\ell+1}\bra{j+1}\nonumber\\
&&+\ket{j}_{\ell}\bra{j+1}\otimes\ket{j+1}_{\ell+1}\bra{j}\bigg]+\textrm{H.C},
\label{Eq.14}
\end{eqnarray}
To achieve this, we consider an eight-component magnetic flux of the form
\begin{equation}
\bar{\Phi}_{\ell,\ell+1}(t)=\sum_{n=1}^{4}\bar{\gamma}_{n}\bigg[\cos(\nu_n t)+\cos(\nu_{n+4} t)\bigg],
\label{Eq.15}
\end{equation}
where the first four frequencies are the same as in the previous case, $\nu_5=\delta_{10}^{10}$, $\nu_6=\delta_{21}^{10}$, $\nu_7=\delta_{21}^{10}$, $\nu_8=\delta_{21}^{21}$ and $\bar{\gamma}_n=\gamma_{n}/2$, with $\gamma_{n}$ defined previously. Under similar conditions, we can perform the RWA and obtain the next effective Hamiltonian
\begin{equation}
H^{(\ell,\ell+1)}_{X}(\mathcal{C}_{x})=\mathcal{C}_{x}S_X^{\ell}S_X^{\ell+1},
\label{Eq.16}
\end{equation}
where $\mathcal{C}_{x}=f\sqrt{Q_AQ_B}$. We can obtain different constants $\mathcal{C}_{xy}$ or $\mathcal{C}_x$ by changing the value of $f$ manipulating the amplitude
$\gamma_{n}$ or $\bar{\gamma}_{n}$ respectively. We can also extend this result for the entire chain using the same magnetic flux (\ref{Eq.15}) through all SQUIDs, thus obtaining in the interaction picture
\begin{equation}
\bar{H}_{X}=\mathcal{C}_{x}\sum_{\ell=1}^{N-1}S_X^{\ell}S_X^{\ell+1}.
\label{Eq.17}
\end{equation}

Therefore, a key ingredient in the protocol is to set the energy differences $\Delta_{kj}^{ml}$ to enable the RWA. As we are interested in simulating a spin-$1$ particle, we forbid the transition $\ket{0}_\ell\rightarrow\ket{2}_\ell$, to do this we choose both values of $g_A$ and $g_B$ in the region where $\ket{0}_{\ell}$ and $\ket{2}_{\ell}$ have the same parity, this happens for $\{g_A,g_B\}>0.5$. Also, we require that $g_A$ and $g_B$ are far enough to obtain appreciable energy differences for the system, obtaining optimal values of $g_A=0.6$ and $g_B=0.9$. Finally, any two-body interaction between sites that are not nearest neighbour is only possible in a dispersive way,  therefore, they are slower and can be neglected. Figure \ref{Fig3N} shows the energy spectrum of Eq. (\ref{Eq.07}) as a function of qubit-resonator coupling $g$, vertical dashed lines indicate the values for the QRS$_A$ ($g_A$) and QRS$_B$ ($g_B$).

Finally, local single-spin rotations can be generated by means of a classical driving $\eta(t)$ acting upon each QRS$_{\ell}$ with
\begin{eqnarray}
&&\eta_{\ell}(t)=\left[\Omega_{\ell}^{(1)}\cos(\mu_{\ell}^{(1)}t+\varphi_{\ell}^{(1)})+\Omega_{\ell}^{(2)}\cos(\mu_{\ell}^{(2)}t+\varphi_{\ell}^{(2)})\right]\left(a_{\ell}^{\dagger}+a_{\ell}\right),\nonumber\\
\label{Eq.18}
\end{eqnarray} 
to see the effect of this driving, we write the interaction Hamiltonian when all signals in the SQUIDs are switched off ($\gamma_n=0$), obtaining for adjacent sites
\begin{eqnarray}
H_D^{\ell,\ell+1}=&&\sum_{\ell=1}^{N}\bigg[\sum_{j,k>j}\bigg(P_{\ell}^{\ell,\ell+1}+P_{\ell}^{\ell-1,\ell}\bigg)z_{kj}^{\ell}e^{i(\epsilon_k^{\ell}-\epsilon_j^{\ell})t}\ket{k}_{\ell}\bra{j}\nonumber\\
&&-2\sqrt{P_{\ell}^{\ell,\ell+1}P_{\ell+1}^{\ell,\ell+1}}\sum_{j,k>j}\sum_{l,m>l}\chi_{kj}^{\ell}\chi_{ml}^{\ell+1}\bigg(e^{i\delta_{kj}^{ml}t}\ket{k}_{\ell}\bra{j}\nonumber\\
&&+e^{i\Delta_{kj}^{ml}t}\ket{j}_{\ell}\bra{k}\bigg)\ket{m}_{\ell+1}\bra{l}+\sum\limits_{l=1}^{2}\frac{\Omega_{\ell}^{(l)}}{2}\sum\limits_{j,k>j}\chi_{kj}^{\ell}\bigg( e^{i[(\epsilon_k^{\ell}-\epsilon_j^{\ell}+\mu_{\ell}^{(l)})t + \varphi_{\ell}^{(l)}]} \nonumber\\
&&+ e^{i[(\epsilon_k^{\ell}-\epsilon_j^{\ell}-\mu_{\ell}^{(l)})t - \varphi_{\ell}^{(l)}]}\bigg)\ket{k}_{\ell}\bra{j}\bigg] + H.C,
\label{Eq.19}
\end{eqnarray}
choosing $\mu_l^{\ell}=\epsilon_{l}^{\ell}-\epsilon_{l-1}^{\ell}$, $\Omega_{\ell}^{(l)}=\sqrt{2}r/\chi_{l,l-1}^{\ell}$, $\varphi_{\ell}^{(l)}=\varphi$ for all $\ell$ and $l$;  with $r$ a manipulable parameter. If $r$ is much smaller than all frequencies in the driving, we can apply the RWA obtaining for one site
\begin{eqnarray}
H_D^{(\ell)}(\varphi)\approx &&\frac{r}{\sqrt{2}}\bigg[ e^{-i\varphi} \bigg(\ket{1}_{\ell}\bra{0} + \ket{2}_{\ell}\bra{1} \bigg) + e^{i\varphi} \bigg( \ket{0}_{\ell}\bra{1}+ \ket{1}_{\ell}\bra{2}\bigg)\nonumber\\
&&=r\bigg(S_X^{\ell}\cos(\varphi)+S_Y^{\ell}\sin(\varphi)\bigg),
\label{Eq.20}
\end{eqnarray}
then, for $\varphi=\pi/2$ and $\varphi=0$ we have
\begin{eqnarray}
H_D^{(\ell)}(\pi/2)=rS_Y^{\ell},\quad H_D^{(\ell)}(0)=rS_X^{\ell},
\label{Eq.21}
\end{eqnarray}
respectively, these interactions produce rotations with respect to the $y$ axis $R_Y^{\ell}e^{-iS_Y^{\ell}\pi/2}$ or $x$ axis $R_Y^{\ell}e^{-iS_Y^{\ell}\pi/2}$ over the $\ell$th QRS. Since all $\eta_{\ell}(t)$ are independent, we can switch on all of them at the same. In particular, for the interacting time $t=\pi/(2r)$, we obtain
\begin{eqnarray}
&&R_X=\prod_{\ell=1}^{N}e^{-iS_X^{\ell}\pi/2},\quad R_Y=\prod_{\ell=1}^{N}e^{-iS_Y^{\ell}\pi/2},
\label{Eq.22}
\end{eqnarray}
which rotate all species simultaneously. This approach allows the generation of simultaneous one and two-body gates between adjacent spin-1 particles defined on each QRS system.


\section{Spin-1 models}
Now we present the protocols for the digital quantum simulation of the Heisenberg model, and for the analog simulation of the Ising model. The anisotropic Heisenberg model of a spin chain of $N$ sites is given by 
\begin{eqnarray}
H_{\textrm{H}}=&&\sum\limits_{\ell=1}^{N-1}\bigg(\lambda_{x}S_{X}^{\ell}S_{X}^{\ell+1} + \lambda_{y}S_{Y}^{\ell}S_{Y}^{\ell+1} + \lambda_{z}S_{Z}^{\ell}S_{Z}^{\ell+1}\bigg)\nonumber\\
=&&\bar{H}_{XY}(\mathcal{C}_{xy})+\bar{H}_{YZ}(\mathcal{C}_{yz})+\bar{H}_{ZX}(\mathcal{C}_{zx})
\label{Eq.23}
\end{eqnarray} 
where
\begin{eqnarray}
\bar{H}_{\alpha\beta}(\mathcal{C}_{\alpha\beta})=&&\mathcal{C}_{\alpha\beta}\sum_{\ell=1}^{N-1}\bigg(S_{\alpha}^{\ell}S_{\alpha}^{\ell+1} + S_{\beta}^{\ell}S_{\beta}^{\ell+1}\bigg)
\label{Eq.24}
\end{eqnarray}
so that $\lambda_{x}=\mathcal{C}_{xy}+\mathcal{C}_{zx}$, $\lambda_{y}=\mathcal{C}_{xy}+\mathcal{C}_{yz}$, and $\lambda_{z}=\mathcal{C}_{yz}+\mathcal{C}_{zx}$,
with $\mathcal{C}_{xy}=f_1\sqrt{Q_AQ_B}$, $\mathcal{C}_{yz}=f_2\sqrt{Q_AQ_B}$, $\mathcal{C}_{zx}=f_3\sqrt{Q_AQ_B}$, where $f_1$, $f_2$ and $f_3$ are different values for $f$ of $\gamma$ using in Eq.~(\ref{Eq.08}). 

\begin{figure}[t]
	\centering
	\includegraphics[width=1\linewidth]{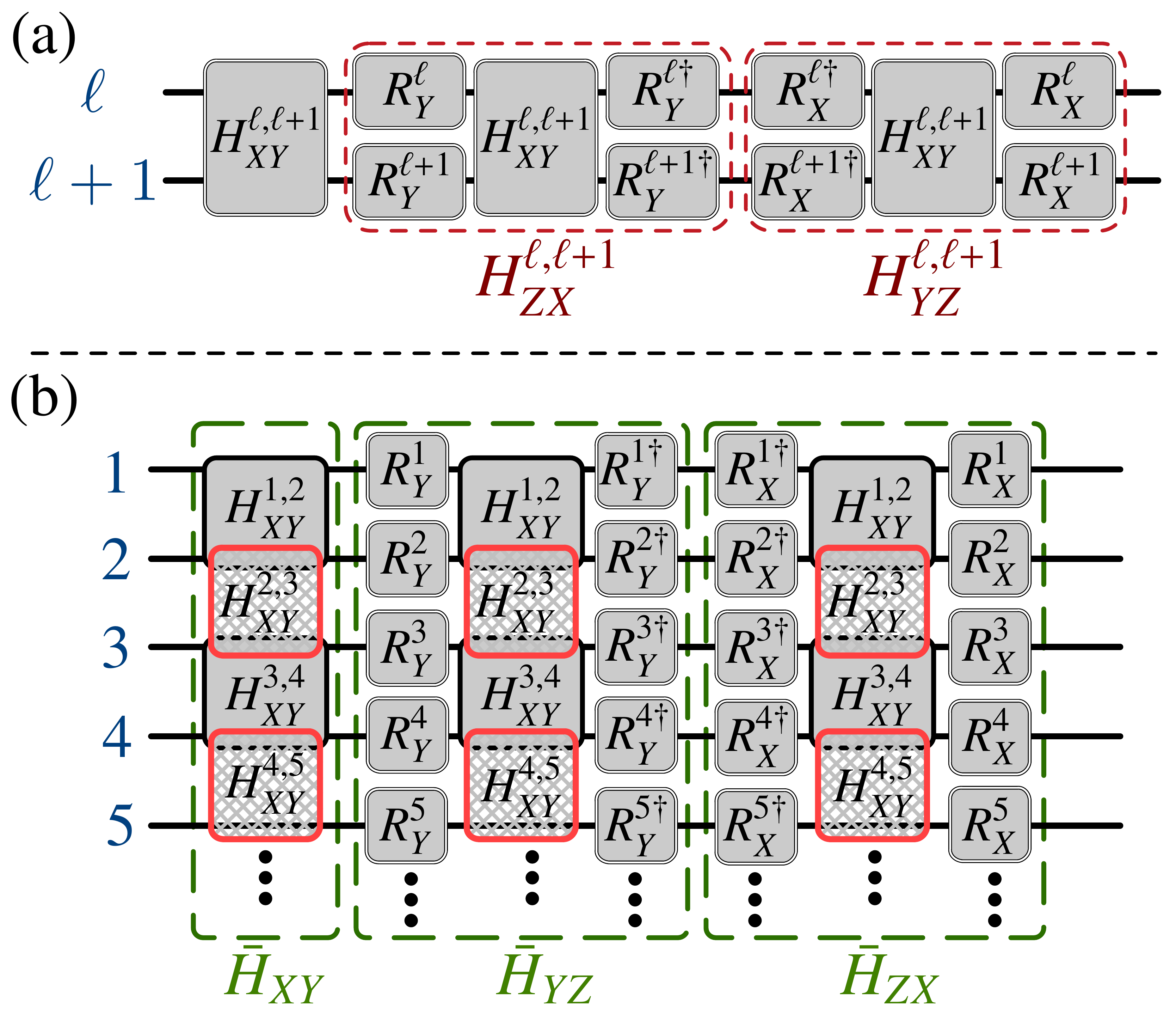}
	\caption{Diagram of the digital quantum simulation of the anisotropic Heisenberg model, for (a) adjacent sites and (b) the entire chain.}
	\label{Fig3}
\end{figure}

The term $\bar{H}_{XY}(\mathcal{C}_{xy})$ in Eq.~(\ref{Eq.23}) is the gate given by Eq.~(\ref{Eq.13}), the terms $\bar{H}_{YZ}(\mathcal{C}_{yz})$ and  $\bar{H}_{ZX}(\mathcal{C}_{zx})$ can be simulated as $R_Y^{\dagger}\bar{H}_{XY}(\mathcal{C}_{yz})R_Y$ and $R_X\bar{H}_{XY}(\mathcal{C}_{zx})R_X^{\dagger}$ respectively, where we use rotations given by Eq.~(\ref{Eq.22}). In this way, the time evolution of the Heisenberg interaction in Eq.~(\ref{Eq.23}) up to time $t$ can be digitally simulated in $n_o$ Trotter steps following the seven steps protocol. \textit{Step 1:} Perform the $R_X^{\dagger}$ rotation. \textit{Step 2:} Evolve the system under the Hamiltonian $\bar{H}_{XY}(\mathcal{C}_{zx})$ for a time $t/n_{o}$. \textit{Step 3:} Perform the $R_X$ rotation. \textit{Step 4:} Perform the $R_Y$ rotation. \textit{Step 5:} Repeat step 2 but with Hamiltonian $\bar{H}_{XY}(\mathcal{C}_{yz})$. \textit{Step 6:} Perform the $R_Y^{\dagger}$ rotation. \textit{Step 7:} Repeat step 2 but with Hamiltonian $\bar{H}_{XY}(\mathcal{C}_{xy})$. This can be summarized as
\begin{eqnarray}
e^{-iH_{\textrm{H}}t}\approx\bigg( e^{-i\bar{H}_{XY}(\mathcal{C}_{xy})t/n_o}R_Y^{\dagger}e^{-i\bar{H}_{XY}(\mathcal{C}_{yz})t/n_o}R_YR_Xe^{-i\bar{H}_{XY}(\mathcal{C}_{zx})t/n_o}R_X^{\dagger}\bigg)^{n_o}.\nonumber\\
\label{Eq.25}
\end{eqnarray}
Figure~\ref{Fig3} (a) shows the gates diagram of this protocol for adjacent sites $(\ell,\ell+1)$, and Fig.~\ref{Fig3} (b) for the entire chain.

$XXZ$ Heisenberg model simulation involves two kind of gates, the $\bar{H}_{XY}$ given by Eq. (\ref{Eq.13}) and the $\bar{H}_{X}$ given by Eq. (\ref{Eq.17}), but switching on all magnetic fields at the same time. The protocol works as follows. \textit{Step 1:} Perform the $R_Y$ rotation. \textit{Step 2:} Evolve the system under Hamiltonian $\bar{H}_{X}(\mathcal{C}_{z})$ for a time $t/n_{o}$. \textit{Step 3:} Perform the $R_Y^{\dagger}$ rotation. \textit{Step 4:} Repeat step 2 but using Hamiltonian $\bar{H}_{XY}(\mathcal{C}_{xy})$. If $\mathcal{C}_{xy}=\mathcal{C}_{z}$ we obtain the isotropic model.

Finally, since we can engineer $\bar{H}_X(\mathcal{C}_X)$ in Eq.~(\ref{Eq.17}), the analog quantum simulation of the Ising model is also feasible via 
\begin{equation}
H_{\textrm{Ising}}=\sum\limits_{\ell=1}^{N-1}JS_{X}^{\ell}S_{X}^{\ell+1} + \sum\limits_{\ell=1}^{N}BS_{X}^{\ell}=\bar{H}_{X}(J)+S_X(B),
\label{Eq.26}
\end{equation}
where the term $S_X(B)=\sum_{\ell=1}^{N}BS_{X}^{\ell}$ can be implemented by a classical driving given by Eq.~(\ref{Eq.15}).


\section{Numerical simulations}
Numerical simulations of spin-1 models are carried out for a chain of $N=4$ sites and in the interaction picture with respect to $H_o=\sum_{\ell=1}^{N}H_o^{\ell}$. We include dissipative effects through the master equation~\cite{Beaudoin2011,Garziano2015,Wu2017}
\begin{eqnarray}
\dot{\rho}=&&-i[H,\rho]+\sum\limits_{\ell}\sum\limits_{j,k>j}\Gamma_{kj}^{(\ell)}[1+\bar{n}(\omega_{kj}^{\ell},T)]\mathcal{D}[\ket{j}_\ell\bra{k}]\rho\nonumber\\
&&+\sum\limits_{\ell}\sum\limits_{j,k>j}\Gamma_{kj}^{(\ell)}[\bar{n}(\omega_{kj}^{\ell},T)]\mathcal{D}[\ket{k}_{\ell}\bra{j}]\rho,
\label{Eq.27}
\end{eqnarray}
where $\mathcal{D}[\hat{O}]\rho=\frac{1}{2}(2\hat{O}\rho\hat{O}^{\dagger}-\rho\hat{O}^{\dagger}\hat{O}-\hat{O}^{\dagger}\hat{O}\rho)$, $\bar{n}$ is the mean photon number of thermal baths at temperature $T$ chosen as $T=15[\rm{mK}]$, and $\omega_{kj}^{\ell}=\epsilon_k^{\ell}-\epsilon_j^{\ell}$. The index $\ell$ stands for the $\ell$th QRS and the frequency-dependent rates have three different component, $\Gamma_{kj}^{(\ell)}=\Gamma_{kj}^{(\ell)\textrm{cav}}+\Gamma_{kj}^{(\ell)\text{dec}}+\Gamma_{kj}^{(\ell)\textrm{deph}}$, where $\Gamma_{kj}^{(\ell)\textrm{cav}}=(\omega_{kj}^{\ell}\kappa_c/\omega^r_\ell)|_\ell\bra{k}(a^{\dagger}_\ell+a_\ell)\ket{j}_{\ell}|^2$ associated to cavity losses, $\Gamma_{kj}^{(\ell)\text{dec}}=(\omega_{kj}^{\ell}\kappa_x/\omega^q_\ell)|_{\ell}\bra{k}\sigma_\ell^x\ket{j}_{\ell}|^2$, associated to qubit decay, and $\Gamma_{kj}^{(\ell)\textrm{deph}}=(\omega_{kj}^{\ell}\kappa_z/\omega^q_\ell)|_{\ell}\bra{k}(\sigma^z_\ell)\ket{j}_{\ell}|^2$ associated to qubit dephasing. We use effective constant in Eq.~(\ref{Eq.01}) $P_{\ell}^{\ell,\ell+1}=Q_{\ell}^{\ell,\ell+1}=3.655[\rm{MHz}]$ \cite{Wang2016} for all $\ell$ and for all simulations (see supplemental material). Also, we use $\kappa_c=2\pi\times10[\rm{kHz}]$, $\kappa_x=2\pi\times20[\rm{kHz}]$ and $\kappa_z=2\pi\times10[\rm{kHz}]$ \cite{LasHeras2014,Wang2016}. Specifically, we carried out the numerical calculations for the digital quantum simulation of the isotropic Heisenberg model, and the analog quantum simulation of the Ising model. In both cases we use a chain of interleaved QRSs of the form $A-B-A-...$, where we fix the parameters of the QRS$_A$ and QRS$_B$ as $\omega_A^r=\omega_B^r=2\pi\times10[\rm{GHz}]$, $\omega_A^q=\omega_B^q=2\pi\times 9[\rm{GHz}]$, $g_A=2\pi\times 6[\rm{GHz}]$ and $g_B=2\pi\times 9[\rm{GHz}]$. Finally, we choose for all cases the SQUID parameter $f=2\pi\times10[\rm{GHz}]$ \cite{Mezzacapo2014b}, which allows to implement an effective coupling between sites $J=f\sqrt{Q_AQ_B}=2\pi\times0.0366[\rm{GHz}]$. We point out that the proposed values of the couplings $g_A$ and $g_B$ are larger than what is nowadays achievable for transmon qubits, while reaching these couplings may require novel circuit designs.

\begin{figure}[t]
	\centering
	\includegraphics[width=1\linewidth]{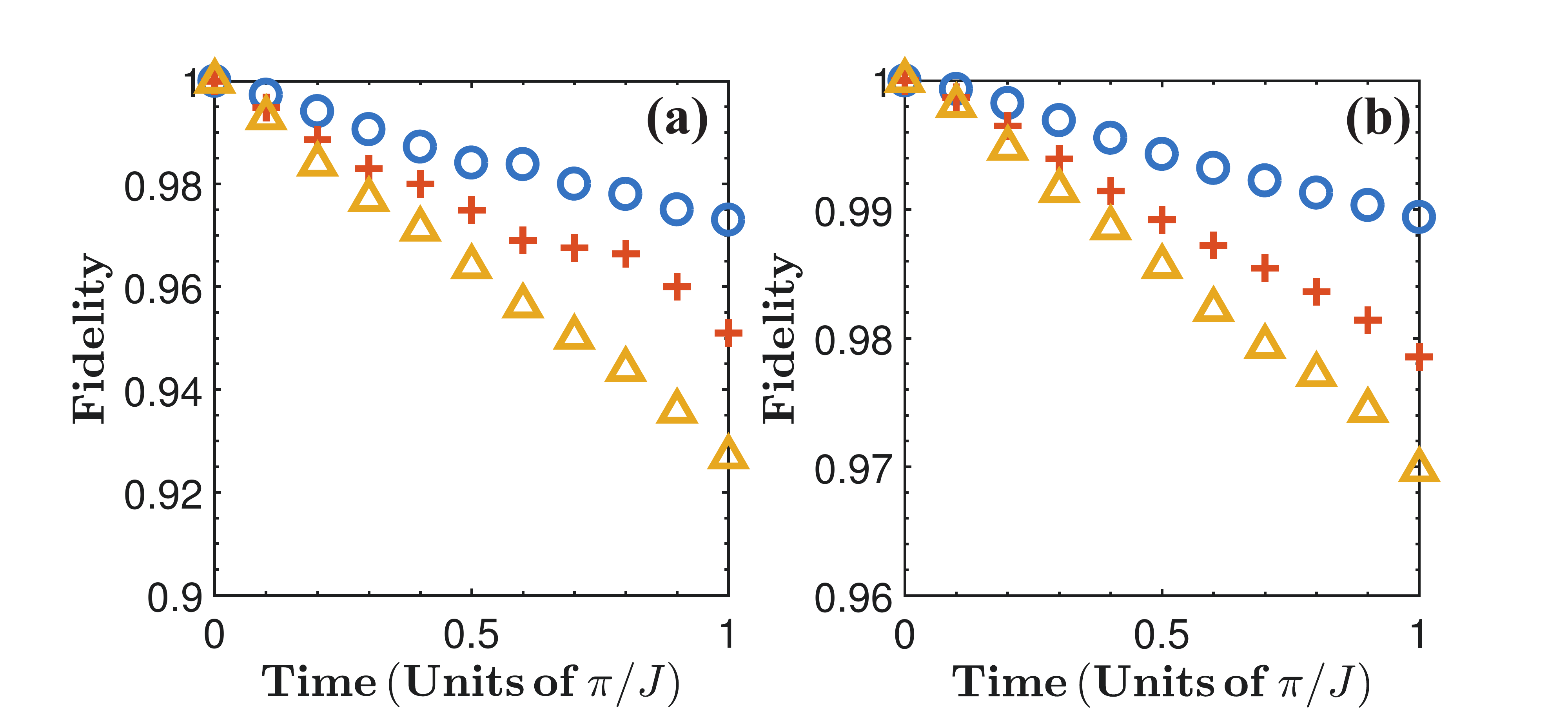}
	\caption{Fidelities as a function of time for the simulation of (a) the anisotropic Heisenberg model and (b) the Ising model. In both figures blue circles stand for a lattice size of $N=2$, red crosses $N=3$ and yellow triangles $N=4$ spin-1 particles.}
	\label{Fig4}
\end{figure}

The simulations are compared with the exact quantum dynamics given by the anisotropic $XXZ$ Heisenberg model, $\lambda_{x}=\lambda_{y}=J$ and $\lambda_{z}=J/2$ in Eq. (\ref{Eq.23})
and the Ising model in Eq.(\ref{Eq.26}). For the latter we choose $B=2\pi\times0.01[\rm{GHz}]$. Figure~\ref{Fig4} shows the average fidelity $F={\rm Tr}\sqrt{\rho^{1/2}\sigma\rho^{1/2}}$ with $\rho$ being the state evolved from the exact Hamiltonian and $\sigma$ the state evolved from the simulated Hamiltonian for $100$ random initial states, being Fig.~\ref{Fig4} (a) for the Heisenberg model and Fig.~\ref{Fig4} (b) for the Ising model. In both figures blue circles stand for a lattice size of $N=2$, red crosses $N=3$ and yellow triangles $N=4$ spin-1 particles. The simulations were done with $n_o=10$ Trotter steps and the elapsed time of the total simulation are $t_{H_2}=t_{H_3}=t_{H_4}\approx0.486[\mu s]$ for the Heisenberg model, and $t_{I_2}=t_{I_3}=t_{I_4}=t=\pi/J\approx0.01[\mu s]$ for the Ising model for $N=2,\,3,\,4$.

The fidelity decreases with the increase of number of particles which is a result of two main sources, that is, the increase of digital errors in the Heisenberg model and the increase of the multi-body gate errors. Nonetheless, the noticeable point of our protocols is that the simulation time does not depend on the number of sites in the chain, such that dissipative processes may have a smaller effect than in usual simulation protocols.

Finally, it is pertinent to mention that though this calculation does not consider multi-mode effects, the main scope of this article would not change, since neither parity nor anharmonicity of the QRM spectrum will be affected if we include multiple resonator modes. Also, the coupling between resonators would not change because they interact via low impedance grounded SQUID, which imposes zero voltage boundary condition at the edge of the resonators. Nevertheless, the multi-mode effects will change the expression for the effective constants $P_{\ell}^{l,l+1}$ and $Q_{\ell}^{l,l+1}$ in Hamiltonian (\ref{Eq.01}), the energy level structure and, therefore, the resonance condition for the activation of specific transitions. A detailed discussion of multi-mode effects in circuit QED has been introduced by A. Parra, \textit{et. al.} in Ref. \cite{Parra2017}. 

\section{Conclusion}
We have proposed a protocol for the digital quantum simulation of spin$-1$ Heisenberg and analog quantum simulation of spin$-1$ Ising models based on a feasible experimental implementation in circuit quantum electrodynamics within the ultrastrong coupling regime. We show how a time-dependent coupling between quantum Rabi systems allows us to activate different two-body transitions without individual QRS manipulation. Finally, we have shown that in our protocols we can activate all one- and two-site interactions at the same time. In this way, we obtain simulation times that are independent of the number of particles in the spin chain, which provides a great potential in scalability. 

F.A.-A. acknowledges support from CONICYT Doctorado Nacional 21140432, G.R. acknowledges funding from FONDECYT under grant No. 1150653, J.C.R.  thanks FONDECYT for support under grant No. 1140194, L.L. acknowledges support from Ram\'on y Cajal Grant RYC-2012-11391, while L.L. and E.S. are grateful for the funding of Spanish MINECO/FEDER FIS2015-69983-P and Basque Government IT986-16.

\appendix
\section{Effective coupling}
\label{AppendixA}
\begin{figure*}[t]
	\centering
	\includegraphics[width=\linewidth]{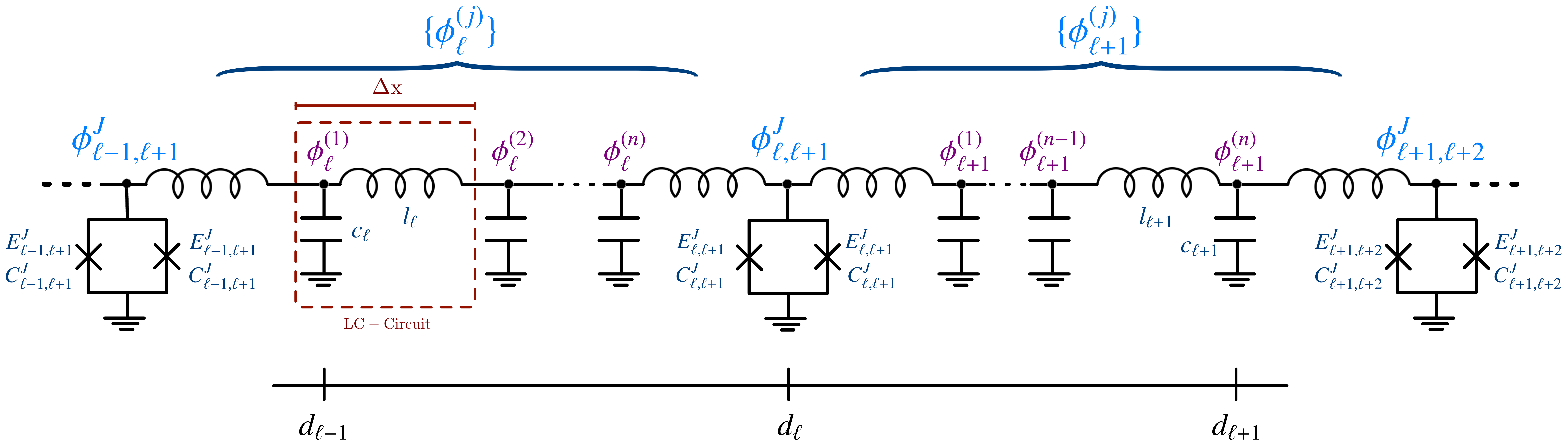}
	\caption{Diagram of the circuit that models two adjacent transmission lines coupled through a grounded SQUID.}
	\label{FigS2}
\end{figure*}

In this section we derive the Hamiltonian of a chain of $N$ transmission lines coupled through SQUIDs, as shown in Fig.~\ref{Fig3N}. We modelled each transmission line (TL) as a set of inductors and capacitors as shown in Fig.~\ref{FigS2} for adjacent sites ~\cite{Felicetti2014}. We use the Hamiltonian circuit description through the spanning tree theory~\cite{Devoret1995}, denoting by $\phi^{(j)}_{\ell}$ the flux associated with the $j$th active node of the $\ell$th transmission line. An inductance per unit length $l_{\ell}$ and a capacitance per unit length $c_{\ell}$ is associated to each resonator. The Lagrangian of the entire chain reads
\begin{equation}
\mathcal{L}=\sum_{\ell=1}^{N}\bigg(\mathcal{L}_{\ell}^{\textrm{TL}}+\mathcal{L}_{\ell,\ell+1}^{\textrm{I}}+\mathcal{L}_{\ell,\ell+1}^{\textrm{S}}\bigg),
\label{A1}
\end{equation}
with
\begin{eqnarray}
\mathcal{L}_{\ell}^{\textrm{TL}}=&&\frac{1}{2}\sum_{j=1}^{n}\left[ c_{\ell}\Delta {\rm x}\left( \dot{\phi}_{\ell}^{(j)}\right) ^2-\frac{1}{l_{\ell}\Delta {\rm x}}\left(\phi^{(j+1)}_{\ell}-\phi^{(j)}_{\ell}\right)^2\right],\nonumber\\
\mathcal{L}_{\ell,\ell+1}^{\textrm{I}}=&&-\frac{1}{2\Delta \rm x}\left[ \frac{1}{l_{\ell}}\left(\phi_{\ell}^{(n)}-\phi^J_{\ell,\ell+1}\right)^2 + \frac{1}{l_{\ell+1}}\left(\phi_{\ell+1}^{(1)}-\phi^J_{\ell,\ell+1}\right)^2 \right], \nonumber\\
\mathcal{L}_{\ell,\ell+1}^{\textrm{S}}=&&\frac{C^J_{\ell,\ell+1}}{2}\left( \dot{\phi}^J_{\ell,\ell+1}\right)^2+2E^J_{\ell,\ell+1}\cos\left( \frac{\Phi^{\rm x}_{\ell,\ell+1}}{2\varphi_o}\right) \cos\left( \frac{\phi^J_{\ell,\ell+1}}{2\varphi_o}\right),\nonumber\\
\label{A2}
\end{eqnarray}
where $\Delta \rm x$ is the characteristic length of each LC circuit, $E^J_{\ell,\ell+1}$ and $\Phi^{\rm x}_{\ell,\ell+1}$ are the Josephson energy and the external magnetic flux threading the SQUID$_{\ell,\ell+1}$ respectively, and we consider symmetric SQUIDs along the chain.  The first equation in (\ref{A2}) corresponds to the $\ell$th TL, the second equation is the interaction between the SQUID and the adjacent TL and, the third term corresponds to the SQUID$_{\ell,\ell+1}$. The Euler-Lagrange (E-L) equation for $\phi_{\ell}^{(j)}$ reads
\begin{equation}
c_{\ell}\Delta{\rm x}\ddot{\phi}_{\ell}^{(j)}=\frac{1}{l_{\ell}\Delta{\rm x}}\bigg[\bigg(\phi_{\ell}^{(j+1)}-\phi_{\ell}^{(j)}\bigg)-\bigg(\phi_{\ell}^{(j)}-\phi_{\ell}^{(j-1)}\bigg)\bigg].
\label{A3}
\end{equation}
The  E-L equation for $\phi^J_{\ell,\ell+1}$ is written as
\begin{eqnarray}
C_{\ell,\ell+1}^J\ddot{\phi}^J_{\ell,\ell+1}=&&\frac{1}{\Delta\rm x}\left[ \frac{1}{l_{\ell+1}}\left( \phi^{(1)}_{\ell+1}-\phi^J_{\ell,\ell+1}\right) + \frac{1}{l_{\ell}}\left( \phi^{(n)}_{\ell}-\phi^J_{\ell,\ell+1}\right)  \right]\nonumber\\
&&-\frac{E_{\ell,\ell+1}^J(\Phi^{\rm x}_{\ell,\ell+1})}{\varphi_o}\sin\bigg(\frac{\phi_{\ell,\ell+1}^J}{2\varphi_o}\bigg),
\label{A4}
\end{eqnarray}
where $E_{\ell,\ell+1}^J(\Phi^{\rm x}_{\ell,\ell+1})=E^J_{\ell,\ell+1}\cos\left( \frac{\Phi^{\rm x}_{\ell,\ell+1}}{2\varphi_o}\right) $. In the continuum limit $\Delta\rm x\rightarrow0$ the Eq.~(\ref{A3}) reads
\begin{equation}
\frac{\partial^2\phi_{\ell}}{\partial x^2}=\frac{1}{\nu^2}\frac{\partial^2\phi_{\ell}}{\partial t^2},
\label{S5}
\end{equation}
with $1/\nu^2=l_{\ell}c_{\ell}$. In the same limit Eq.~(\ref{A4}) reads
\begin{eqnarray}
&&\frac{1}{l_{\ell+1}} \left( \frac{\partial}{\partial {x}}\phi_{\ell+1}\right)\biggr\rvert_{{\rm x}=d_{\ell}} -\frac{1}{l_{\ell}}\left( \frac{\partial}{\partial {x} }\phi_{\ell}\right)\biggr\rvert_{{\rm x}=d_{\ell}}\nonumber\\
&&=\frac{E^J_{\ell,\ell+1}(\Phi^{\rm x}_{\ell,\ell+1})}{\varphi_o}\sin\left(  \frac{\phi(d_{\ell},t)}{2\varphi_o}\right)+C^J_{\ell,\ell+1}\ddot{\phi}(d_{\ell},t).
\label{A6}
\end{eqnarray}

Equation~(\ref{A3}) is the wave equation, which can be solved using separation of variables $\phi_{\ell}(x,t)=\chi_{\ell}(x)\tau_{\ell}(t)$, thus, for the spatial function we obtain
\begin{equation}
\frac{d^2}{dx^2}\chi_{\ell}=-(\bar{\kappa}_{\ell})^2\chi_{\ell}\Rightarrow\chi_{\ell}(x)=A\sin(\bar{\kappa}_{\ell} x) + B\cos (\bar{\kappa}_{\ell} x),
\label{S7}
\end{equation}
In the low impedance limit of the SQUIDs,  the boundary conditions for each bulk resonator are $\chi(d_{\ell})=0$, for $\ell=\{1,2,...,N-1\}$. The boundary conditions for the edges of the chain establish that no current is flowing which means $\chi'(0)=\chi'(d_{N})=0$. Then, the spatial function reads
\begin{equation}
\chi_{\ell}(x)=A_{\ell}^n\sin[\bar{\kappa}_{\ell}^n (x-d_{\ell-1})], 
\label{A8}
\end{equation}
where $\bar{\kappa}_{\ell}^n=\frac{n\pi}{d_{\ell}-d_{\ell-1}}$, for $\ell=\{2,3,...,N-1\}$, $\bar{\kappa}_{1}^n=\frac{\pi}{d_{1}}(n+1/2)$ and $\bar{\kappa}_{N}^n=\frac{\pi(n+1/2)}{d_{N}-d_{N-1}}$. These conditions define a $\lambda/4$-resonator for the edges of the chain and $\lambda/2$-resonators for the rest of the chain.

Assuming that for each SQUID the Josephson energy is much larger than the charging energy, that is the phase regime, we can approximate $\sin(\phi(d_{\ell},t)/(2\varphi_o))\approx\phi(d_{\ell},t)/(2\varphi_o)$, $\cos(\phi(d_{\ell},t)/(2\varphi_o))\approx1-\phi(d_{\ell},t)^2/(8\varphi_o^2)$. Also, if the plasma frequencies of SQUIDs are the largest scales compared with low-lying frequencies in the system, we can neglect the last term of Eq.~(\ref{A6}) since we can consider the system dynamics slower \cite{Johansson2009}. These approximations lead to
\begin{equation}
\frac{E^J_{\ell,\ell+1}(\Phi^{\rm x}_{\ell,\ell+1})}{2\varphi_o^2}\phi(d_{\ell},t)=\frac{1}{l_{\ell+1}}\left( \frac{\partial}{\partial x}\phi_{\ell+1}\right)\biggr\rvert_{x=d_{\ell}}-\frac{1}{l_{\ell}}\left( \frac{\partial}{\partial x}\phi_{\ell}\right)\biggr\rvert_{x=d_{\ell}}.
\label{A9}
\end{equation}
To calculate the Hamiltonian, we integrate the spatial modes of the Lagrangian (\ref{A1}) in the continuum limit, obtaining for the $\ell$th transmission line
\begin{equation}
\int\mathcal{L}^{\textrm{TL}}_{\ell}dx=\left( \frac{C_{\ell}}{2}\dot{\tau}^{n2}_{\ell}-\frac{\kappa_{\ell}^{n2}}{2L_{\ell}}\tau_{\ell}^{n2}\right)
,\quad\kappa_{\ell}^n=\bar{\kappa}_{\ell}^n(d_{\ell}-d_{\ell-1}).
\label{A10}
\end{equation}
The Lagrangian for the SQUIDs, the last term of Eq.~(\ref{A1}), in the harmonic approximation reads
\begin{equation}
\mathcal{L}^{S}_{\ell,\ell+1}=-E^J_{\ell,\ell+1}(\Phi^{\rm x}_{\ell,\ell+1})\frac{\phi(d_{\ell},t)^2}{4\varphi_o^2},
\label{A11}
\end{equation}
and using the condition of Eq.~(\ref{A9}) for $\phi(0,t)$, we obtain
\begin{equation}
\mathcal{L}_{\ell,\ell+1}^{S}=-\frac{\varphi_o^2}{E^J_{\ell,\ell+1}(\Phi^{\rm x}_{\ell,\ell+1})}\left( \frac{\kappa^n_{\ell+1}}{L_{\ell+1}}\tau^n_{\ell+1} - \frac{\kappa^n_{\ell}}{L_{\ell}}\tau^n_{\ell}\right)^2,
\label{A12}
\end{equation}
then, the total Lagrangian for the lowest modes of each resonator reads
\begin{eqnarray}
\mathcal{L}_{sys}=\int\mathcal{L}dx=&&\sum_{\ell=1}^N\left( \frac{C_{\ell}}{2}\dot{\tau}_{\ell}^{2}-\frac{\kappa_{\ell}^{2}}{2L_{\ell}}\tau_{\ell}^{2}\right)\nonumber\\
&&-\frac{\varphi_o^2}{E^J_{\ell,\ell+1}(\Phi^{\rm x}_{\ell,\ell+1})}\left( \frac{\kappa_{\ell+1}}{L_{\ell+1}}\tau_{\ell+1} - \frac{\kappa_{\ell}}{L_{\ell}}\tau_{\ell}\right)^2.\quad
\label{A13}
\end{eqnarray}
Now, using the canonical conjugate variable $p_{\ell}=\partial\mathcal{L}/\partial\dot{\tau}_{\ell}=C_{\ell}\dot{\tau}_{\ell}$, we can write the Hamiltonian as 
\begin{eqnarray}
\mathcal{H}=&&\sum_{\ell=1}^{N}p_{\ell}\dot{\tau}_{\ell}-\mathcal{L}_{sys}=\sum_{\ell=1}^N\Bigg( \frac{p_{\ell}^2}{2C_{\ell}}+\frac{\kappa_{\ell}^{2}}{2L_{\ell}}\tau_{\ell}^{2}\Bigg)\nonumber\\
&&+\sum_{\ell=1}^{N-1}\frac{\varphi_o^2}{E_{\ell,\ell+1}^{J}(\Phi^{\rm x}_{\ell,\ell+1})}\left( \frac{\kappa_{\ell}}{L_{\ell}}\tau_{\ell} - \frac{\kappa_{\ell+1}}{L_{\ell+1}}\tau_{\ell+1}\right)^2,
\label{A14}
\end{eqnarray}
defining $\omega_{\ell}=\kappa_{\ell}/\sqrt{C_{\ell}L_{\ell}}$, we obtain
\begin{equation}
\mathcal{H}=\sum_{\ell=1}^N\left( \frac{p_{\ell}^{2}}{2C_{\ell}}+\frac{C_{\ell}}{2}\omega_{\ell}^2\tau_{\ell}^2\right)+\frac{\varphi_o^2}{E_{\ell,\ell+1}^J(\Phi^{\rm x}_{\ell,\ell+1})}\left( \frac{\omega_{\ell}}{Z_{\ell}}\tau_{\ell} - \frac{\omega_{\ell+1}}{Z_{\ell+1}}\tau_{\ell+1}\right)^2,
\label{A15}
\end{equation}
where $Z_{\ell}=\sqrt{L_{\ell}/C_{\ell}}$. Using the standard quantization procedure
\begin{eqnarray}
p_{\ell}=i\sqrt{\frac{\hbar C_{\ell}\omega_{\ell}}{2}}(a_{\ell}^{\dagger}-a_{\ell}), \quad \tau_{\ell}=\sqrt{\frac{\hbar}{2C_{\ell}\omega_{\ell}}}(a_{\ell}^{\dagger}+a_{\ell}),\qquad
\label{A16}
\end{eqnarray}
the Hamiltonian reads
\begin{eqnarray}
\hat{H}=&&\sum_{\ell=1}^N\hbar\omega_{\ell}a_{\ell}^{\dagger}a_{\ell}+\sum_{\ell=1}^{N-1}\frac{\hbar\varphi_o^2}{2E^J_{\ell,\ell+1}(\Phi^{\rm x}_{\ell,\ell+1})}\bigg[ \frac{1}{Z_{\ell}}\sqrt{\frac{\omega_{\ell}}{C_{\ell}}}(a_{\ell}^{\dagger}+a_{\ell})\nonumber\\
&&- \frac{1}{Z_{\ell+1}}\sqrt{\frac{\omega_{\ell+1}}{C_{\ell+1}}}(a_{\ell+1}^{\dagger}+a_{\ell+1})\bigg]^2.
\label{A17}
\end{eqnarray}
Now, we assume that all SQUIDs are equal, then $E_{\ell,\ell+1}^J=E_J$. We consider the external flux $\Phi^{\rm x}_{\ell,\ell+1}$ to be composed by a DC signal and a small AC signal as $\Phi^{\rm x}_{\ell,\ell+1}=\Phi^o_{\ell,\ell+1}+\Phi_{\ell,\ell+1}(t)$. Since $|\Phi_{\ell,\ell+1}(t)|\ll\Phi^o_{\ell,\ell+1}$, we can expand $1/E_J(\Phi^{\rm x}_{\ell,\ell+1})$ as
\begin{eqnarray}
\frac{1}{E_J(\Phi^{\rm x}_{\ell,\ell+1})}=&&\frac{1}{E_J\cos\left(\frac{\Phi^o_{\ell,\ell+1}+\Phi_{\ell,\ell+1}(t)}{2\varphi_o} \right) }
\nonumber\\
\approx&& \frac{1}{\bar{E}_J}\bigg(1 +\frac{\sin(\bar{\Phi}^o_{\ell,\ell+1})}{\cos(\bar{\Phi}^o_{\ell,\ell+1})}\bar{\Phi}_{\ell,\ell+1}(t)\bigg)
\label{A18}
\end{eqnarray}
where $\bar{E}_J=E_J\cos(\bar{\Phi}^o_{\ell,\ell+1})$, $\bar{\Phi}^o_{\ell,\ell+1}=\frac{\Phi^o_{\ell,\ell+1}}{2\varphi_o}$ and $\bar{\Phi}_{\ell,\ell+1}(t)=\frac{\Phi(t)_{\ell,\ell+1}}{2\varphi_o}$. We can rewrite the Hamiltonian (\ref{A17}) as
\begin{eqnarray}
\hat{H}&&=\sum_{\ell=1}^N\bigg[ \hbar\omega_{\ell}a_{\ell}^{\dagger}a_{\ell}+\hbar \bigg(P_{\ell}^{\ell,\ell+1}+P_{\ell}^{\ell-1,\ell}\bigg)(a^{\dagger}_{\ell}+a_{\ell})^2\Bigg] \nonumber\\
&&-\sum_{\ell=1}^{N-1}\bigg[2\sqrt{P_{\ell}^{\ell,\ell+1} P_{\ell+1}^{\ell,\ell+1}}(a_{\ell}^{\dagger}+a_{\ell})(a_{\ell+1}^{\dagger}+a_{\ell+1})\bigg]\nonumber\\
&&+\sum_{\ell=1}^{N}\bigg[\bigg(Q_{\ell}^{\ell,\ell+1}\bar{\Phi}_{\ell,\ell+1}(t)+Q_{\ell}^{\ell,\ell-1}\bar{\Phi}_{\ell,\ell-1}(t)\bigg)(a^{\dagger}_{\ell}+a_{\ell})^2\bigg]\nonumber\\
&&-\sum_{\ell=1}^{N-1}\bigg[2\sqrt{Q_{\ell}^{\ell,\ell+1} Q_{\ell+1}^{\ell,\ell+1}}\bar{\Phi}_{\ell,\ell+1}(t)(a_{\ell}^{\dagger}+a_{\ell})(a_{\ell+1}^{\dagger}+a_{\ell+1})\bigg],\quad\qquad
\label{A19}
\end{eqnarray}
where
\begin{eqnarray}
&&P_{\ell}^{l,l+1}=\frac{\varphi_o\omega_{\ell}}{4I_c Z_{\ell}^{2}C_{\ell}}\frac{1}{\cos(\bar{\Phi}_o^{l,l+1})},\nonumber\\
&&Q_{\ell}^{l,l+1}=\frac{\varphi_o\omega_{\ell}}{4I_c Z^{2}_{\ell}C_{\ell}}\frac{\sin(\bar{\Phi}_o^{l,l+1})}{\cos^2(\bar{\Phi}_o^{l,l+1})},
\label{A20}
\end{eqnarray}
with $I_c=E_J/(2\varphi_o)$ the critical current. For the simulation in the main text we use for sites $\ell$, the flux quantum $\varphi_o=3.2911[\rm{fWb}]$, the critical current $I_c=1[\rm{mA}]$, the transmission line impedance $Z_{\ell}=Z=100[\Omega]$ and capacitance $C_{\ell}=C=200[\rm{fF}]$, and same time independent offset component of the external magnetic flux $\bar{\Phi}_o^{l,l+1}=\bar{\Phi}_o=\pi/4$ for all SQUIDS; given us effective constants $P_{\ell}^{l,l+1}=P=Q_{\ell}^{l,l+1}=Q=3.655[\rm{MHz}]$.    

%


\end{document}